\documentstyle[11pt]{article}

\def \Rem       {{\rm Re}_{\rm m}}

\title{On the Kinematic Dynamo Action by ABC Flows} 

\author{S.\ B.\ F.\ Dorch\thanks{E-mail address: dorch@astro.su.se} \\
      The Royal Swedish Academy of Sciences,\\
      Stockholm Observatory,\\ 
      SE-133 36 Saltsj\"{o}baden, Sweden } 

\date{PACS: 52.30.-q, 52.30.Jb, 
            95.30.Qd, 96.50.Bh, 96.60.Qc, 97.10.Jb }

\begin{document}

\maketitle

\begin{abstract}
The kinematic induction equation of MHD is solved numerically 
using a staggered mesh in the case of 
the normal `111' ABC flow. Basic integral properties such as the growth rate
of magnetic energy agree with previous findings, but careful 3D visualizations
of the topology of the magnetic field
reveal, however, that the conclusions about the
modes of operation of this kinematic dynamo must be revised. 
The two known windows of dynamo action at low and high magnetic Reynolds number,
correspond to two distinct modes, both relying on the replenishing of the
magnetic field near the $\beta$ type stagnation points in the flow. 
The results support the case for the normal ABC flow as a fast dynamo.
\end{abstract}

\section{Introduction}

The class of steady velocity flows called ABC flows 
(after Arnold, Beltrami, and Childress) is an example of 
how complex flows may be able to amplify weak seed magnetic fields. 
ABC flows have been studied in e.g.\ the context of the cyclonic 
$\alpha$-effect by \cite{Childress1970}, 
but the main reason that this rather special type of 
flow is interesting to dynamo theory, 
is that it on the one hand constitutes 
a situation so simple that it is possible to understand the details of the 
amplification process, and that it one the other hand displays a behavior 
that in some respect resembles that of astrophysical dynamos such 
as the Sun: 
The amplification is related to the stretching, twisting and folding of 
magnetic field lines, and
the r\^{o}le played by the finite diffusivity is
as essential to the amplification process as it is believed to be for the
Sun. The main purpose of this paper is to illustrate the principle of 
operation of a dynamo by means of visualizing the 3-D structure of magnetic
field lines and their dynamics. Doing so we discover yet unknown properties
of this type of dynamo.
  
\section{ABC Flows}

The ABC flow is a 3-D periodic, incompressible and steady flow given by
the sum of three parameterized Beltrami waves
\cite{Beltrami1889}: 
\begin{equation}
{\bf u} = A(0,\sin kx, \cos kx)+B(\cos ky,0,\sin ky)+C(\sin kz,\cos
kz,0). \label{ABCflow}
\end{equation}
Traditionally the approach is purely kinematic, so that the back-reaction in the 
equation of motion by the Lorentz force is ignored in the treatment
(but see \cite{Galanti+ea92}).  
Thus the problem reduces to one of solving
the induction equation,
\begin{equation}
	\frac{\partial {\bf B}}{\partial t} = - \nabla\times ({\bf u}\times 
	{\bf B}) + \eta \nabla^2 {\bf B}, \label{induct}
\end{equation}
where ${\bf B}$ is the magnetic field,
${\bf u}$ is the velocity field (Eq.\ \ref{ABCflow})
and $\eta$ is the magnetic diffusivity.

\subsection{Method}

The induction equation is solved with
a simplified version of the code 
by Galsgaard, Nordlund and others (e.g.\
\cite{Galsgaard+Nordlund97} and \cite{Nordlund+ea94}),
using both uniform magnetic resistivity 
and the quenched diffusion that is normally used in the
full version of this MHD code:
In both cases the magnetic Reynold number $\Rem$ increases with the 
size of the grid as $n^2$ (where the number of grid points is $n^3$).
In this paper only results from cases with constant and uniform  
diffusion are presented.
 
The induction equation is solved on a periodic finite difference staggered
mesh with 6th order staggering operators, 5th order centering operators,
and a 3rd order Hyman time stepping routine.

Even though the flow is steady, in certain 
regions of the flow trace particles follow
chaotic paths so that any two trace particles separate
exponentially with time --- a property that has been shown to 
be essential for dynamo action (\cite{Ott93}).
By varying the basic parameters of
the flow $A$, $B$, $C$, and the wavenumber $k$ it turns out that a lot
of both quantitatively as well as qualitatively different topologies are 
possible: A special case with $A=0$ is the G.\ O.\ Roberts dynamo used in 
the context of the Earth's dynamo (\cite{Rotvig1998}).
While the {\em normal} ABC flow with $A=B=C=1$ 
(hence also called the ``111'' flow) has 8 stagnation
points, a flow with $A=5$ and $B=C=2$ has no stagnation points.
If all the parameters are non-zero the flow in general contains
a mixture of chaotic and regular islands (\cite{STF}).

The normal ABC flow has two regimes of dynamo action: One at comparatively
low magnetic Reynolds numbers $\Rem$ = 8.5 -- 17.5 
(\cite{Arnold+ea83}) 
and one starting at $\Rem$ $\approx$ 27 
(\cite{Galloway+ea84}
and \cite{Galloway+ea86}).

To initialize the mode in the high $\Rem$ regime one may use
a weak uniform magnetic seed field pointing in any direction. 
By flux conservation the uniform part remains constant, and thus
rapidly becomes insignificant relative to the growing mode.
To initialize the growing mode in the low $\Rem$ window, one may use the 
following 
periodic initial condition 
(see \cite{STF} and \cite{Galloway+ea92}),
\begin{equation}
  {\bf B} = (\sin (kz) - \cos (ky), 
             \sin (kx) - \cos (kz), 
             \sin (ky) - \cos (kx)), \label{special}
\end{equation}
which is an eigenmode of Eq.\ \ref{induct} in the case of zero
diffusivity $\eta = 0$ and is related to the rate of strain matrix
$\partial u_i / \partial x_j$ of the ABC flow.

\subsection{Topology}

The normal ``111'' ABC flow has 8 stagnation points and 8 points where the
velocity is maximum (see Figure \ref{figure1}).
A stagnation point is a point where the fluid velocity vanishes, and since the
vorticity is aligned with the velocity vectors for Beltrami waves 
(\cite{Milne-Thomson1955}), 
the vorticity also vanishes at stagnation points. 
There are two different types of stagnation points for the normal ABC flow:
\begin{itemize}
	\item $\alpha$ type stagnation points where stream lines are 
diverging along an axis through the stagnation
point and converging in the plane perpendicular to the axis.
	\item $\beta$ type stagnation point where stream lines are 
converging along the axis and diverging in the plane.
\end{itemize}
The stream lines of the flow have a three-fold symmetry in the converging 
(diverging) planes through the $\alpha$ ($\beta$) type 
stagnation points (see Figure \ref{figure1}). The three-fold symmetric
`leaves' of the converging/diverging stream lines are separated by 
convergence/divergence connect to other stagnation points: 
While the separator lines and leaves of diverging stream 
lines of a $\beta$ point connect to three $\alpha$ type stagnation points the 
reverse is true for an $\alpha$ point. 

In itself it is not so important that there happens to be 8 stagnation points 
in the normal ABC flow. After all, many interesting flows do not possess 
stagnation points (such e.g.\ the  flow, see
\cite{Galloway+Proctor92} and turbulent flows in general),
and in any case a stagnation point may be removed by a simple translation of 
the coordinate system. Rather it is the stretching ability of the flow that
is relevant for the dynamo action.  
For a divergence free flow $\nabla \cdot {\bf u} = 0$ such as the
ABC flow the induction equation may be written in the form 
\begin{equation}
\label{bstretch}
\frac{\partial {\bf B}}{\partial t} + {\bf u}\cdot\nabla{\bf B} =
{\bf B}\cdot\nabla {\bf u} + \eta\nabla^2 {\bf B}
\end{equation}
Apart from diffusive effects the stretching term ${\bf B}\cdot\nabla {\bf u}$ 
implies a linear growth of the logarithm of the magnetic field magnitude.

In the high degree of symmetry of the normal ABC flow the stagnation points
coincide with local extrema of the stretching rate. Hence the two types of 
stagnation points may be seen as convenient markers of these regions.  

\section{Results}

Below the results of experiments with normal ABC flows are divided
into results concerning the growth of e.g.\ magnetic energy and exponential
growth rates and results
related to the formation and evolution of magnetic structures.

\subsection{Magnetic energy}

As shown in Figure \ref{figure2}, the initially weak seed field is 
indeed amplified in an exponential manner, and the growth increases with
$\Rem$. For intermediate $\Rem$ (in the second window), an oscillating 
behavior is associated with the growth (as also seen by e.g.\ 
\cite{Galloway+ea86}, \cite{Galanti+ea92} 
and \cite{Galanti+ea93}), see 
Figure \ref{figure2}.
The oscillation may be understood
as a direct consequence of the spatial periodicity of the ABC flow 
(\cite{Galanti+ea92}).
The period of the oscillation increases with increasing $\Rem$ until a transition
to a non-oscillating regime occurs
at a $\Rem$ of about 200 
(\cite{Lau+Finn93}). Figure \ref{figure2} shows
examples from all three $\Rem$ regimes.

As mentioned by \cite{Galloway+ea92} and 
\cite{STF} it is possible to rig
the initial condition of the seed magnetic field such that only one mode is
present in the calculations; the special initial condition for which that is
possible is given by Eq.\ \ref{special}.

In that case the magnetic field is not amplified in the second window
as mentioned by \cite{Galloway+ea92}. 
\cite{STF} used this initial condition together with the so called
`flux conjecture' to try to deduce the limiting growth rate of the ABC flow.
They found exponential growth of the flux in a selected region.  The growth
rate derived does not, however, agree with the asymptotic growth rates found
by \cite{Galloway+ea92} and \cite{Lau+Finn93}. One should indeed 
not expect to be able to recover the growth rate of a completely different
(exponential growing) mode by studying the stretching of field lines
in a another (secularly decaying) mode.

Figure \ref{figure3} shows that the mode in the case of the initial 
condition given by Eq.\ \ref{special} is an exponentially decaying mode:
The decay is associated with an oscillation with a period about 10 times smaller
than that of the growing mode. 
Because of numerical round-off errors the amplitude of the growing mode
is not identically equal to zero in the initial condition, and eventually
its inevitable growth and the decay of the initialized mode results in a
transition from decay to growth of the total magnetic energy (see 
Figure \ref{figure3}).
\cite{Galloway+ea92} found no growing solution and they
concluded that ``something odd is going on''. 

\subsection{Flux cigars}

When the induction equation is evolved from a weak, uniform seed field,
flux ``cigars'' rapidly arise at four of the eight stagnation points --- 
the $\alpha$-points (see Figure \ref{figure4}). These are the regions 
in the flow where the magnetic field is most rapidly advected by a converging flow in
two of the three dimensions. The flux cigars are aligned along
the axis of divergence through the $\alpha$ type stagnation points and
point directly to the $\beta$ type stagnation points. 
The regions around the $\beta$ type stagnation points are not
unstable to flux cigar generation since the flow there is diverging in two
directions.  However, the converging flow along the
axis from the $\alpha$ points creates weaker, sheet like flux concentrations
on both sides of the plane of divergence (see Figure 
\ref{figure8}).

The four flux cigars seen in Figure \ref{figure4} correspond to the fastest
growing eigenmode in the diffusion-less case $\eta =0$. Initially several
modes may be present but eventually the one with the largest exponential
growth rate will be the only one that remains in the solution.

In our experiments we do indeed find that a 
second set of flux cigars is formed next to the 4 primary cigars, but 
we conclude that rather than obscuring the physics, they are absolutely
essential. 
These secondary cigars have the opposite polarity of the 
neighboring cigars, and they are connected by reconnecting field lines 
(see Figure \ref{figure5}).

The secondary flux cigars arise after the initial transient phase.  
These cigars begin to form at the $\alpha$ point at the separator line next 
to the primary cigar. As the secondary cigar forms the primary cigar
moves slightly away from the center of the stagnation point. 
In the simulation within the oscillatory regime, as the energy 
increases at the beginning of 
a `cycle', the secondary cigar increases in size and field strength until 
the two cigars become equal in size midway through the cycle 
(Figure \ref{figure6}). The primary cigar
now becomes smaller and actually vanishes at the end of the cycle, at which
time the growth of the energy slows down, and even reverses briefly. 
At that point the flux cigar that was
previously the secondary cigar moves into the center of the stagnation
point (see the evolution in Figure \ref{figure6}).

\subsection{A dynamical amplification process}

The dynamical picture 
of an amplification cycle may be sketched as follows in the 
intermediate-$\Rem$ oscillating regime (see e.g.\ the case of $\Rem$ = 40 in
Figure \ref{figure2}): 

In the beginning of a cycle, at one of the local minima in the
magnetic energy, there is only one flux cigar at each $\alpha$ point
(see Figure \ref{figure6}). 
Gradually field lines of the opposite polarity pile up at the separator line 
close to the $\alpha$ point where a secondary cigar will eventually form. 
These field lines come from the plane of divergence of a neighboring $\beta$ 
point and are transported along two
sets of converging stream lines on opposite sides of the separator line.
The stream lines come from two leaves of diverging stream lines at the 
$\beta$ point separated by the separator line: Where the stream lines reach the
$\alpha$ point they have twisted so that the field lines they carry 
are parallel to the axis of divergence through the $\alpha$ point. 
At the other side of the $\alpha$ point one of the leaves of diverging stream 
lines from a 
$\beta$ point supply field lines of the opposite polarity to the primary flux
cigar. While these two flux cigars with opposite polarity sit at the
stagnation point and receive field lines, the field lines that they are made 
of may reconnect because of their opposite polarities 
(see in Figure \ref{figure5}),
and the non-vanishing 
magnetic diffusivity. This reconnection process is essential to the operation 
of the dynamo. If there were no reconnections of field lines, the magnetic
field near the $\beta$ points could 
not be replenished.  It is replenished by tight field 
line `hooks' that are released to move out along the axis of divergence through 
the $\alpha$ point towards the plane of the $\beta$ point.
The field lines of the legs of these hooks that come from the primary cigar
are stretched out into 
triangular shapes by the three-fold symmetric diverging stream lines in the 
plane of a $\beta$ point (see Figure \ref{figure7}). 
The field lines in this leg of the hook go through the plane of divergence of
the $\beta$ point and connect the two primary cigars on each side of 
the plane.
The part of the field line that makes up the bases of the 
triangle line up perpendicular to the separator lines and are added to the two
flux cigars at the $\alpha$, points at the end of these lines. The corners 
of the triangle are also added to $\alpha$ points but through the diverging
leaves of the $\beta$ point, that become the converging leaves of three other
$\alpha$ points. 
In the plane of the $\beta$ points an intricate folding of field lines
takes place: Figure \ref{figure8} shows isosurfaces of weak magnetic field
and field lines near the plane of a $\beta$ point. The plane actually 
constitutes a region of discontinuity where the field lines change direction
from above to below.

Towards the end of a cycle, the primary cigar begins to decay,
because 
the supply of flux to the primary cigar is less than that to the
secondary cigar. At the end of one cycle the secondary cigar is the only
remaining one and it moves to the center of the stagnation point.

The magnetic energy grows through most of the cycle but the main growth takes
place while the two cigars have about equal sizes. This is the time at
which the most rapid reconnection take place and thus the time where the largest
amount of magnetic flux is
released down along the axis of divergence through the 
$\alpha$ points.

The cycle discussed above is actually only a half cycle: A full cycle
requires that the flux cigars return to the original polarity and this is 
only achieved after two of the above cycles. In this respect the ABC dynamo
is similar
to the solar dynamo where the full cycle of the dynamo takes two 11 year 
migrations of the sunspot pattern before the polarity returns to the original.

For large $\Rem$ (in the non-oscillatory regime) the
amplification process is
similar to what was described above for the
lower $\Rem$ case. However, for large $\Rem$, there are no oscillation 
associated with the amplification process: The double cigars always consist of
one large strong cigar and one smaller and weaker one with the opposite 
polarity. The secondary cigars never becomes stronger than the primary but
the mechanism that drives the amplification process remains the same.

In the case of the single-cigar 
mode in the low $\Rem$ window, the mode pulsates with a high frequency,
corresponding roughly to the time of transport of field in a logical
circle.  Magnetic field of the opposite polarity to that of a pre-existing
cigar is transported towards the $\alpha$ point by the three `leaves' 
of converging flows.  
The three leaves do not provide equal amounts of flux but because the 
diffusion is relatively large this does not result in the formation of
a discrete cigar of opposite polarity.  Rather, the existing cigar
is surrounded from all sides by magnetic flux of opposite polarity
(cf.\ Figure \ref{figure9}),
its field is cancelled, and a flux cigar of opposite polarity
is formed.  As a result of the polarity inversion, the polarity of
the magnetic flux ejected from the $\beta$ point region is reversed,
and the cycle repeats.

An analysis of work and dissipation shows that the average work done
against the Lorentz force is positive throughout the cycle but that
the average Joule dissipation is smaller than the average work in the
growing phase of the cycle and larger than the average work in the
decaying phase of the cycle.  Averaged over a cycle, the work is
slightly larger than the dissipation, and hence the mode energy grows.

\section{Discussion and conclusions}

The dynamo action in the two windows of $\Rem$ correspond to two distinct 
modes. In both cases the replenishing of the field near the $\beta$ points
is crucial for the operation of the dynamo.
In these regions a complicated folding of field lines
takes place, that is accompanied by a discontinuity of the directions of field
lines across the plane. The latter result would have been hard to discover
without the use of modern visualization technics that allow direct 
``eye-contact'' when data browsing. 
 
Certain properties of the magnetic transport are nearly invariant as
$\Rem$ is increased:
The size of the regions where diffusion is important become smaller as
$\Rem^{-\frac{1}{2}}$ but  reconnection still takes place and the 
field in the crucial $\beta$ regions continues to be replenished.
In the bulk of the flow where the stretching takes place the decrease of 
the diffusivity is unimportant since the field lines there are not influenced
by diffusion: They tend to obtain a certain alignment with the flow 
topology given by the stretching that is an invariant property
and hence the linear rate of increase of $\ln B$ remains nearly the same.
It thus appears very unlikely that the double-cigar mode should go away
in the limit of infinite $\Rem$ and that the normal 111 ABC flow would not be
a fast dynamo.

\begin{flushleft}
{\bf Acknowledgements}\\
The author acknowledges support through an EC-TMR grant to the European Solar 
Magnetometry Network and thanks V.\ D.\ Archontis and {\AA}.\ Nordlund for
help and discussions.
\end{flushleft}

\newpage

\begin{figure}
\caption[]{A view showing the structure of the normal ABC flow:
Stagnation points are marked with purple isosurfaces, regions of maximum
speed with blue isosurfaces, and stream lines are white. Near the center of the 
view is a $\beta$ type stagnation point in a plane of three-fold symmetry with 
diverging stream lines. In the upper right corner is an $\alpha$ type stagnation
point connecting to the $\beta$ type point via 
a heteroclinic orbit; along this line the velocity increases from zero to
a maximum half-way, decreasing again towards zero at the $\beta$ point.}
\label{figure1}
\end{figure}

\begin{figure}
\caption[]{Four panel showing the magnetic energy as function
of time for the cases: $\Rem$ = 12, 40, 200, and 1600. } 
\label{figure2}
\end{figure}

\begin{figure}
\caption[]{The total magnetic energy $E_M$ as a function of time
on a logarithmic scale for two experiments (moderate $\Rem$) with uniform
initial condition (thick line) and the initial condition given by
Eq.\ \ref{special} (thin line).} 
\label{figure3}
\end{figure}

\begin{figure}
\caption[]{A snapshot of magnetic field strength isosurfaces at a high
value (blue). Also shown are stagnation points (green) and regions of 
maximum speed (purple).
A blue flux cigar is 
at a green ($\alpha$ type) stagnation point in the upper corner (just above
the center of the picture).
This cigar points towards a purple region of maximum speed, a green stagnation
point ($\beta$ type), yet another purple maximum speed region, and in the
bottom corner is seen the opposite (periodic)
end of the flux cigar.} 
\label{figure4}
\end{figure}

\begin{figure}
\caption[]{A snapshot from a high $\Rem$ experiments
showing the structure of the double flux cigars (transparent
purple) located at an $\alpha$ type stagnation point, also shown are magnetic 
field lines in white. } \label{figure5}
\end{figure}

\begin{figure}
\caption[]{Four snapshots (only part of the box is shown) of an
experiment with a moderate $\Rem$ showing the evolution of the double flux 
cigars (yellow). Also shown are converging stream lines in white and the 
diverging axis through the $\alpha$ point (in red). The four pictures correspond
to four instants during the oscillation in the magnetic energy, at fractions
of 0, 0.14, 0.87 and 1 of the cycle.} \label{figure6}
\end{figure}

\begin{figure}
\caption[]{A snapshot from a high $\Rem$ experiment
showing double flux cigars (purple) and field lines. The black and red
field lines are connected in a `hook' between the double cigars in the closer
corner. The field lines
are stretched out in a triangle by the diverging (blue) stream lines in the
plane of the $\beta$ point at the center of the view. } \label{figure7}
\end{figure}

\begin{figure}
\caption[]{A snapshot of weak magnetic field isosurfaces and
field lines around a $\beta$ type stagnation point. The vertical field lines
near the top of the picture come from the double cigars at an $\alpha$ point.}
\label{figure8}
\end{figure}

\begin{figure}
\caption[]{A snapshot from an experiments with $\Rem$=12,
and the initial condition for the magnetic field given by Eq.\ 
\ref{special} showing magnetic field lines (two sets with opposite
polarity) and isosurfaces, at a time near a local minimum
of magnetic energy.  The flux cigar in the center is surrounded
by magnetic field of opposite polarity, and Joule dissipation
is high between the two polarities.} \label{figure9}
\end{figure}

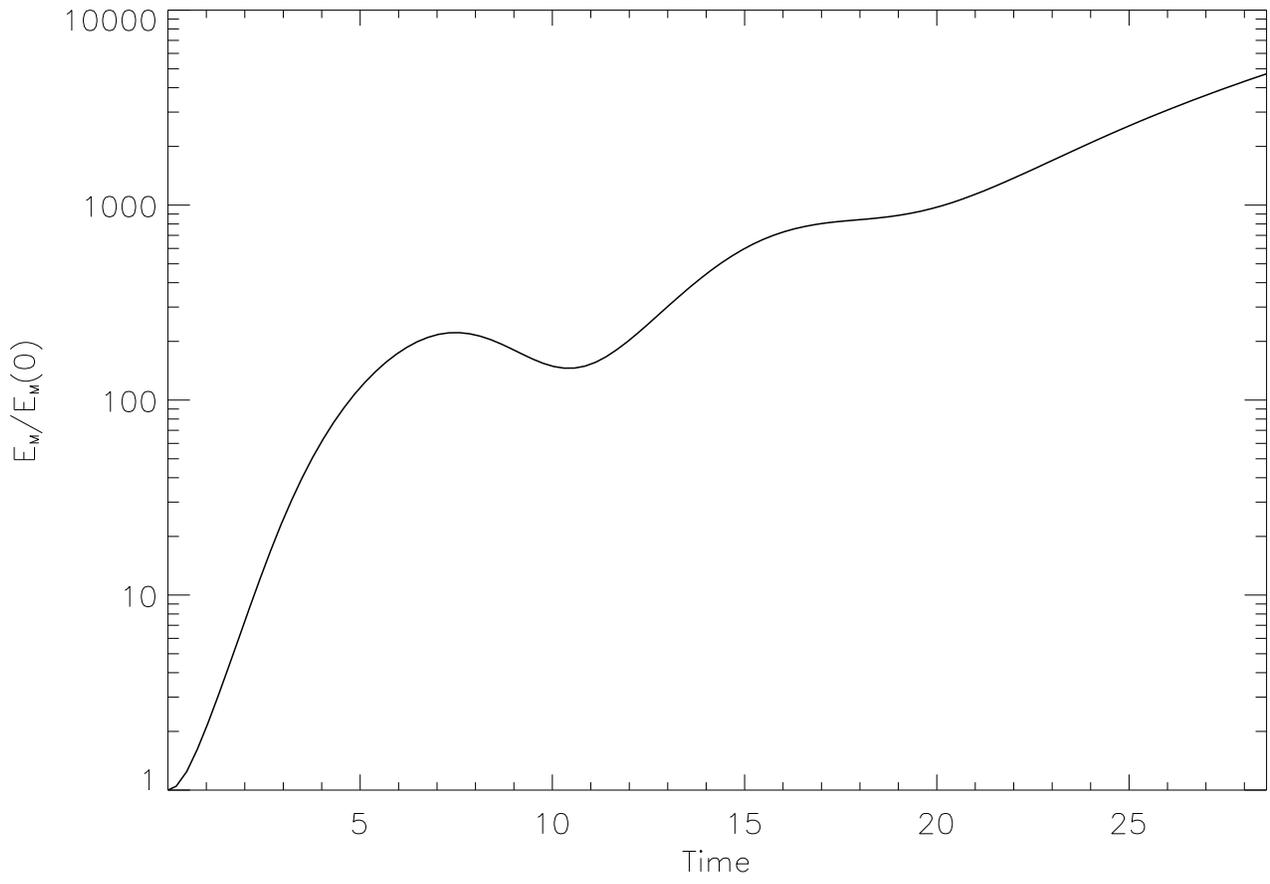
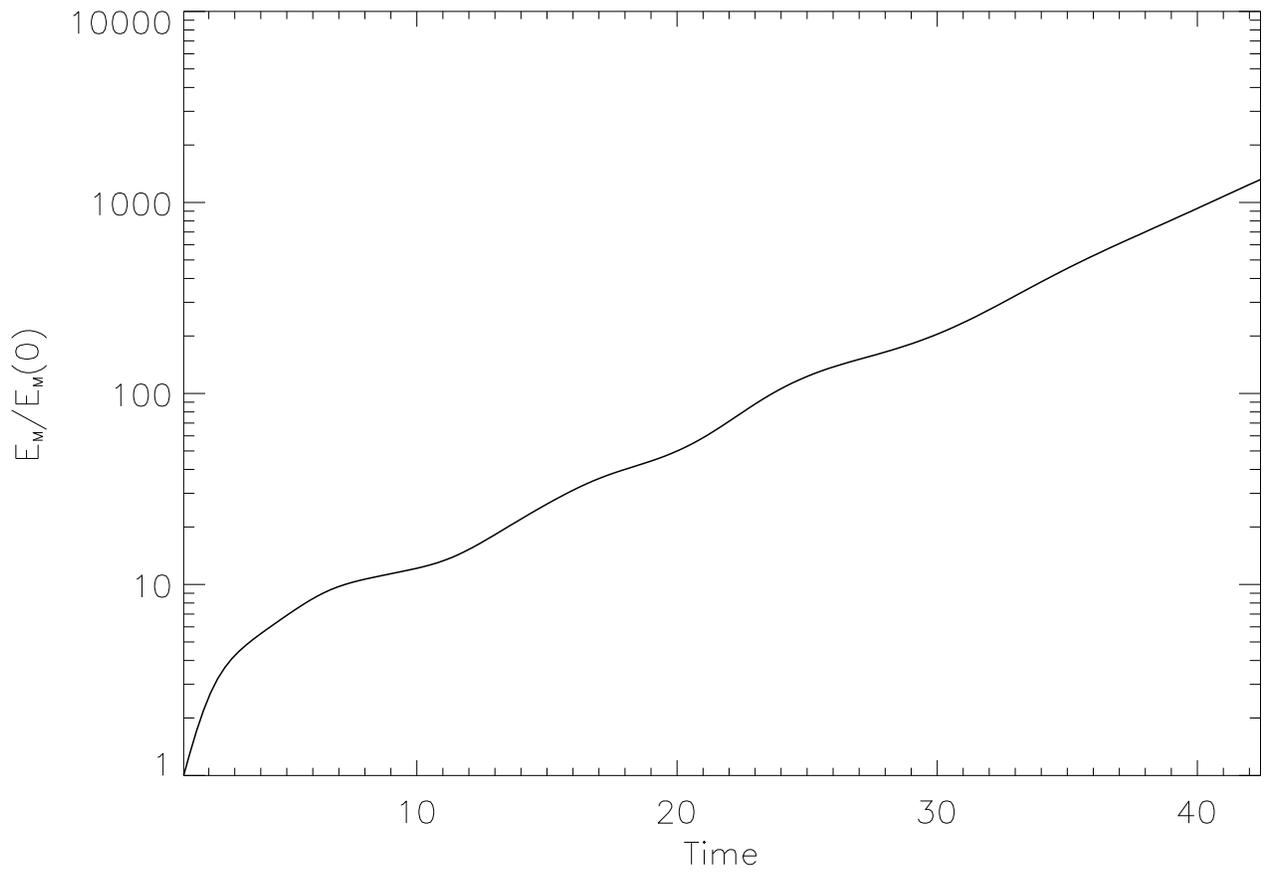
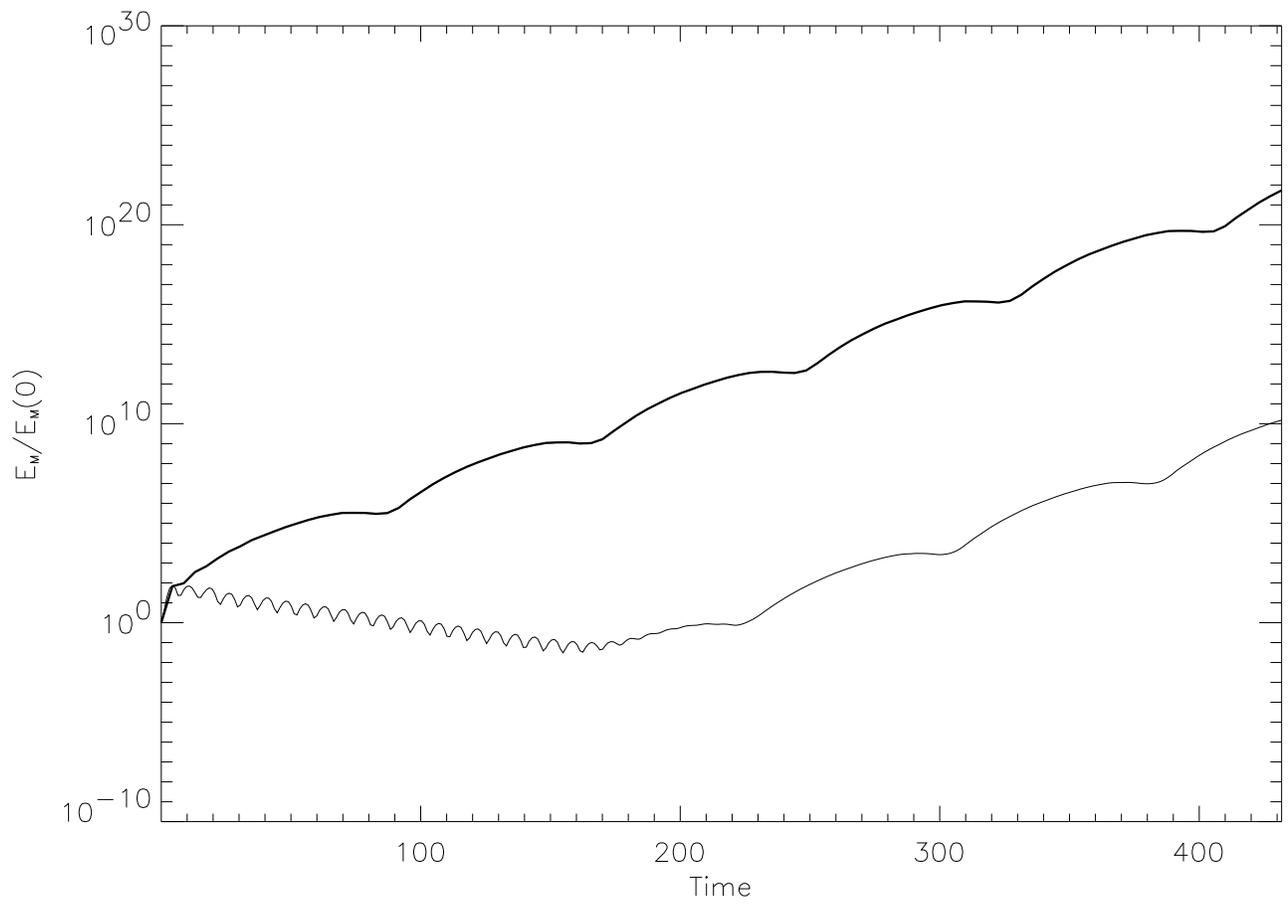
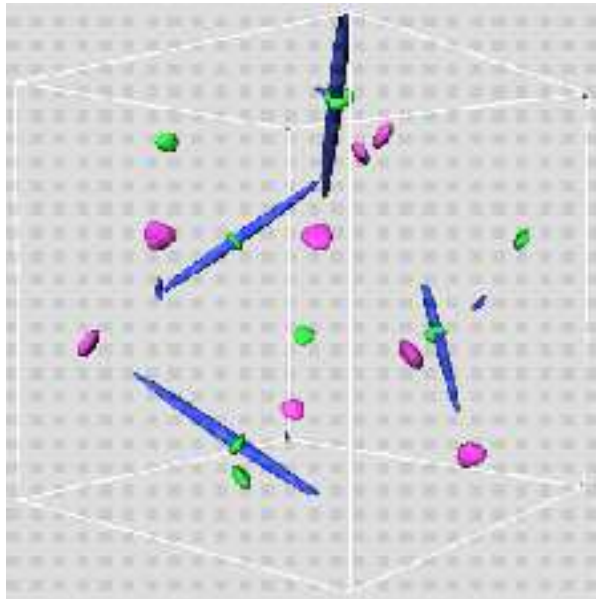
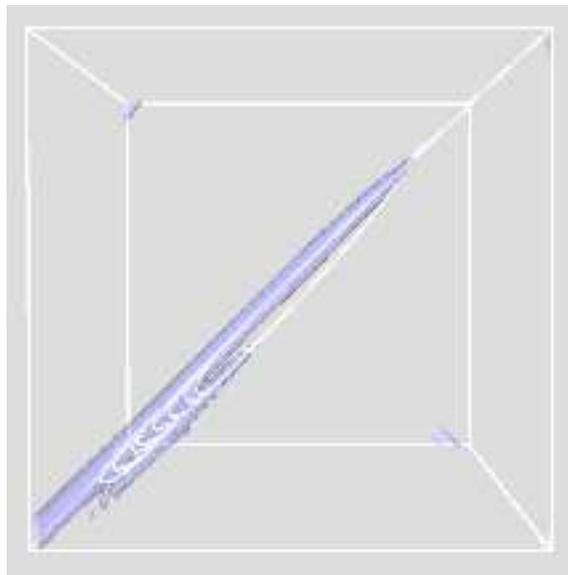
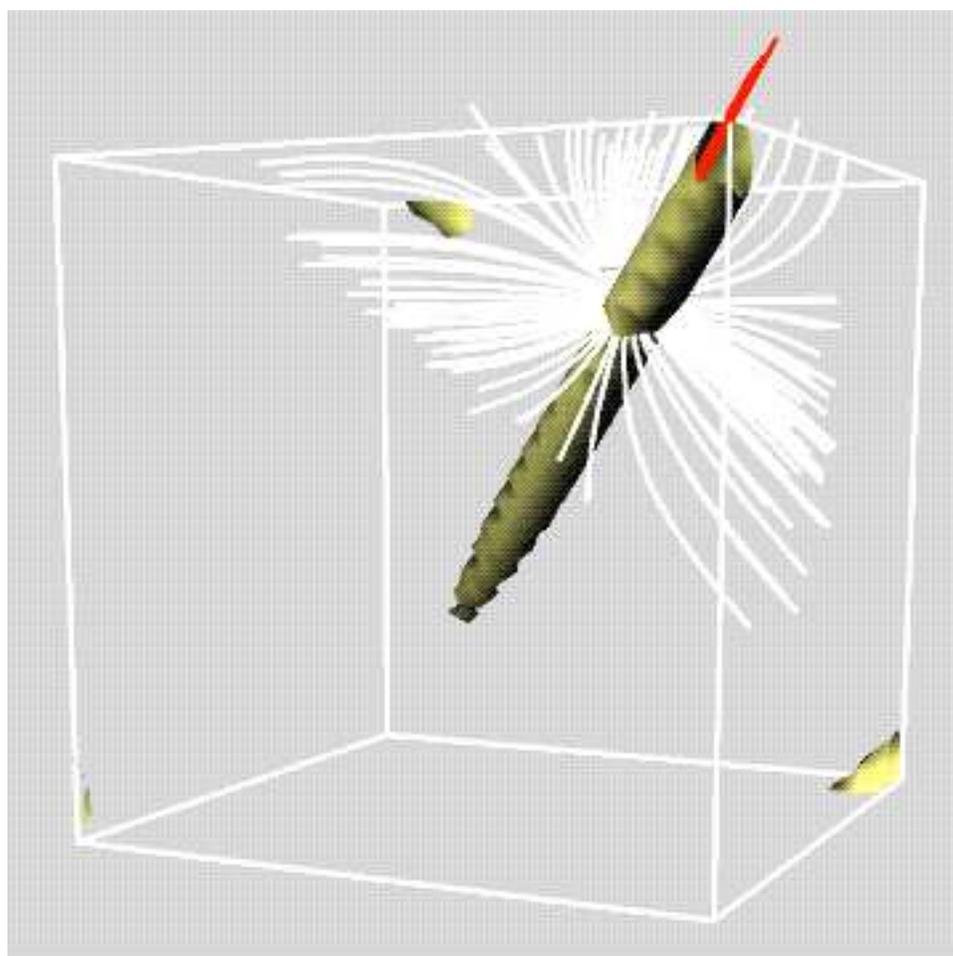
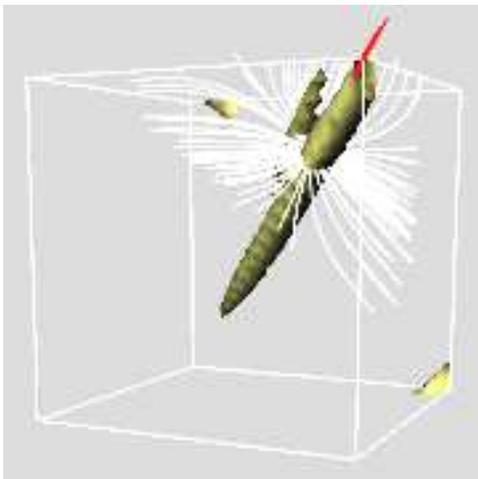
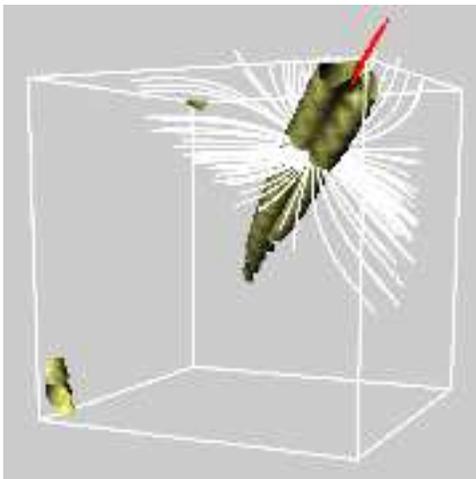
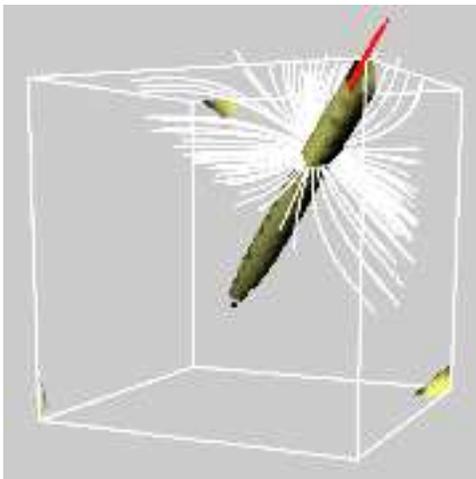
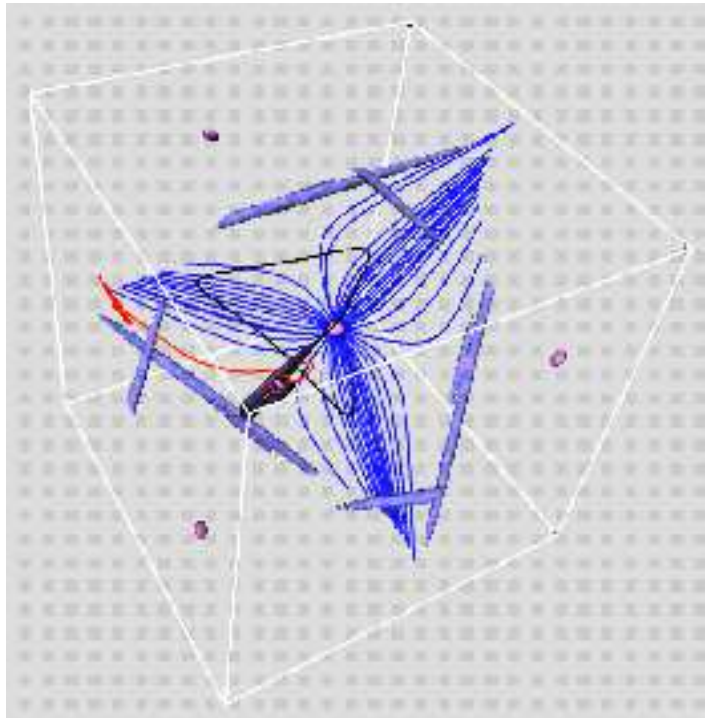
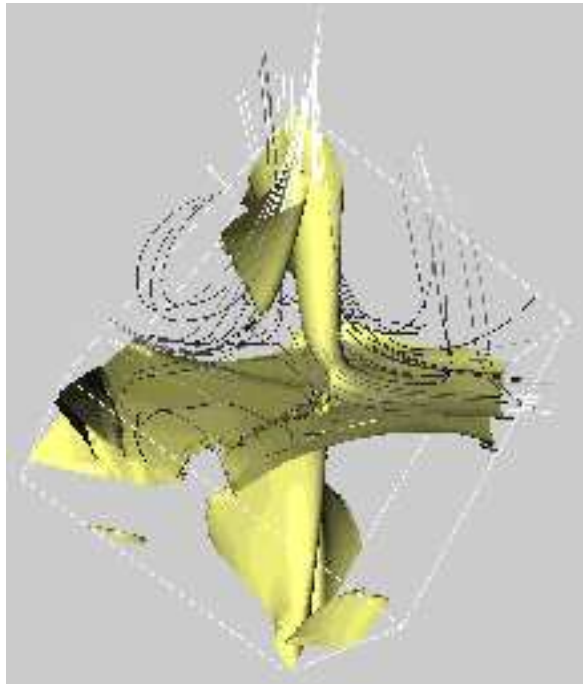
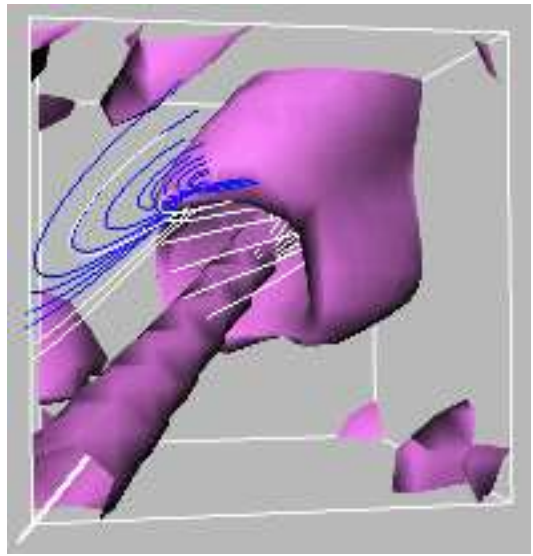
\end{document}